\documentclass[usenatbib]{mn2e}
\usepackage{graphicx,times}
\usepackage{bm,url}
\newcommand{\EQ}{\begin{equation}}
\newcommand{\EN}{\end{equation}}
\newcommand{\EQA}{\begin{eqnarray}}
\newcommand{\ENA}{\end{eqnarray}}

\newcommand{\EEqs}[2]{Equations~(\ref{#1}) and~(\ref{#2})}
\newcommand{\Eq}[1]{equation~(\ref{#1})}
\newcommand{\Eqs}[2]{equations~(\ref{#1}) and~(\ref{#2})}

\newcommand{\App}[1]{Appendix~\ref{#1}}
\newcommand{\Sec}[1]{Sect.~\ref{#1}}

\newcommand{\Fig}[1]{Fig.~\ref{#1}}

\newcommand{\bra}[1]{\langle #1\rangle}

{}
{}
\newcommand{\meanFFFF}{\overline{\mbox{\boldmath ${\cal F}$}}{}}{}

{}
{}
{}
\newcommand{\meanEMF}{\overline{\mbox{\boldmath ${\cal E}$}}{}}{}
{}
{}
{}
{}
\newcommand{\meanAA}{\overline{\mbox{\boldmath $A$}}{}}{}
\newcommand{\meanBB}{\overline{\mbox{\boldmath $B$}}{}}{}
\newcommand{\meanFF}{\overline{\mbox{\boldmath $F$}}{}}{}
{}
{}
{}
\newcommand{\meanJJ}{\overline{\mbox{\boldmath $J$}}{}}{}
\newcommand{\meanUU}{\overline{\mbox{\boldmath $U$}}{}}{}
\newcommand{\meanWW}{\overline{\mbox{\boldmath $W$}}{}}{}
{}
\newcommand{\meanQQ}{\overline{\mbox{\boldmath $Q$}}{}}{}

\newcommand{\meanB}{\overline{B}}

\newcommand{\meanU}{\overline{U}}

\newcommand{\meanJ}{\overline{J}}

\newcommand{\meanW}{\overline{W}}

{}

{}

{}
{}

\newcommand{\ww}{\mbox{\boldmath $w$} {}}

\newcommand{\uu}{\mbox{\boldmath $u$} {}}
\newcommand{\UU}{\mbox{\boldmath $U$} {}}

\newcommand{\aaa}{\mbox{\boldmath $a$} {}}
\newcommand{\bb}{\mbox{\boldmath $b$} {}}
\newcommand{\BB}{\mbox{\boldmath $B$} {}}

\newcommand{\jj}{\mbox{\boldmath $j$} {}}
\newcommand{\JJ}{\mbox{\boldmath $J$} {}}
\newcommand{\SSS}{{\sf S}}
\newcommand{\AAA}{\mbox{\boldmath $A$} {}}

\newcommand{\ff}{\mbox{\boldmath $f$} {}}

\newcommand{\FF}{\mbox{\boldmath $F$} {}}

\newcommand{\WW}{\mbox{\boldmath $W$} {}}

\newcommand{\nab}{\mbox{\boldmath $\nabla$} {}}

\newcommand{\oo}{\mbox{\boldmath $\omega$} {}}

\newcommand{\ggamma}{\mbox{\boldmath $\gamma$} {}}

\newcommand{\SSSS}{\mbox{\boldmath ${\sf S}$} {}}

\newcommand{\DD}{{\rm D} {}}

\newcommand{\dd}{{\rm d} {}}
\newcommand{\const}{{\rm const}  {}}

\def\la{\mathrel{\mathchoice {\vcenter{\offinterlineskip\halign{\hfil
$\displaystyle##$\hfil\cr<\cr\sim\cr}}}
{\vcenter{\offinterlineskip\halign{\hfil$\textstyle##$\hfil\cr<\cr\sim\cr}}}
{\vcenter{\offinterlineskip\halign{\hfil$\scriptstyle##$\hfil\cr<\cr\sim\cr}}}
{\vcenter{\offinterlineskip\halign{\hfil$\scriptscriptstyle##$\hfil\cr<\cr\sim\cr}}}}}
\def\ga{\mathrel{\mathchoice {\vcenter{\offinterlineskip\halign{\hfil
$\displaystyle##$\hfil\cr>\cr\sim\cr}}}
{\vcenter{\offinterlineskip\halign{\hfil$\textstyle##$\hfil\cr>\cr\sim\cr}}}
{\vcenter{\offinterlineskip\halign{\hfil$\scriptstyle##$\hfil\cr>\cr\sim\cr}}}
{\vcenter{\offinterlineskip\halign{\hfil$\scriptscriptstyle##$\hfil\cr>\cr\sim\cr}}}}}
\def\Ma{\mbox{\rm Ma}}

\def\Pm{P_{\rm m}}

\def\Rm{R_{\rm m}}

\def\cs{c_{\rm s}}
\def\kf{k_{\rm f}}
\def\urms{u_{\rm rms}}

\def\nut{\nu_{\rm t}}
\def\nuT{\nu_{\rm T}}
\def\etat{\eta_{\rm t}}

\def\etaT{\eta_{\rm T}}
\def\mut{\mu_{\rm t}}
\def\muT{\mu_{\rm T}}

\def\half{{\textstyle{1\over2}}}

\def\onethird{{\textstyle{1\over3}}}

\def\twothird{{\textstyle{2\over3}}}

\newcommand{\yapj}[3]{ #1, {ApJ,} {#2}, #3}

\newcommand{\yapjl}[3]{ #1, {ApJ,} {#2}, #3}

\newcommand{\yan}[3]{ #1, {Astron.\ Nachr.,} {#2}, #3}

\newcommand{\yana}[3]{ #1, {A\&A,} {#2}, #3}

\newcommand{\ygafd}[3]{ #1, {Geophys.\ Astrophys.\ Fluid Dyn.,} {#2}, #3}

\newcommand{\yjfm}[3]{ #1, {J.\ Fluid Mech.,} {#2}, #3}

\newcommand{\ypfb}[3]{ #1, {Phys.\ Fluids B,} {#2}, #3}

\newcommand{\yprl}[3]{ #1, {Phys.\ Rev.\ Lett.,} {#2}, #3}

\newcommand{\ymn}[3]{ #1, {MNRAS,} {#2}, #3}
\newcommand{\ynat}[3]{ #1, {Nature,} {#2}, #3}

\newcommand{\ysph}[3]{ #1, {Solar Phys.,} {#2}, #3}

\newcommand{\ypre}[3]{ #1, {Phys.\ Rev.\ E,} {#2}, #3}

\newcommand{\ybook}[3]{ #1, {#2} (#3)}

\title[Yoshizawa effect and Archontis dynamo]%
{The role of the Yoshizawa effect in the Archontis dynamo}
\author[S.~Sur and A.~Brandenburg]%
{Sharanya Sur$^{1}$ and Axel Brandenburg$^{2,3}$
\thanks{E-mail: sur@iucaa.ernet.in (SS); brandenb@nordita.org (AB)}\\
$^{1}$Inter-University Centre for Astronomy and
        Astrophysics,  Post Bag 4, Ganeshkhind, Pune 411 007, India\\
$^2$NORDITA, AlbaNova University Center, Roslagstullsbacken 23,
SE 10691 Stockholm, Sweden\\
$^3$Department of Astronomy, AlbaNova University Center,
Stockholm University, SE 10691 Stockholm, Sweden
}

\date{}
\begin{document}

\pagerange{\pageref{firstpage}--\pageref{lastpage}} \pubyear{2009}

\maketitle

\begin{abstract}
The generation of mean magnetic fields is studied for a simple non-helical
flow where a net cross helicity of either sign can emerge.
This flow, which is also known as the Archontis flow, is a generalization
of the Arnold--Beltrami--Childress flow, but with the cosine terms omitted.
The presence of cross helicity leads to a mean-field dynamo effect
that is known as the Yoshizawa effect.
Direct numerical simulations of such flows demonstrate the presence of
magnetic fields on scales larger than the scale of the flow. 
Contrary to earlier expectations, the Yoshizawa effect is found to be
proportional to the mean magnetic field and can therefore lead to its
exponential instead of just linear amplification for magnetic
Reynolds numbers that exceed a certain critical value.
Unlike $\alpha$ effect dynamos, it is found that the Yoshizawa effect is
not noticeably constrained by the presence of a conservation law.
It is argued that this is due to the presence of a forcing term in the
momentum equation which leads to a nonzero correlation with the magnetic field.
Finally, the application to energy convergence in solar wind turbulence
is discussed.

\end{abstract}
\label{firstpage}

\section{Introduction}

The dynamo effect in astrophysical objects is often associated with the
occurrence of helicity in them.
In magnetohydrodynamics there are several helicities that can be important.
A particularly important one is the kinetic helicity, because its
value is finite in rotating stratified bodies and can lead to an
$\alpha$ effect \citep{Mof78,Par79,KR80}.
Another important helicity is the magnetic helicity.
Unlike the kinetic helicity, the magnetic helicity is conserved by the
quadratic interactions, so its value can only change through
resistive effects or through magnetic helicity fluxes \citep{BS05}.
Such a conservation law is crucial to understanding the
saturation behavior of $\alpha$ effect dynamos.
This is because the $\alpha$ effect tends to produce large-scale
magnetic fields that are helical, but conservation of total magnetic
helicity implies that there must be small-scale magnetic helicity
of the opposite sign, so that the sum of small-scale and large-scale
magnetic helicities is close to zero.
This then leads to a resistively slow saturation phase in the nonlinear
regime \citep{B01}.
Mathematically, the consequence of magnetic helicity conservation can be
described by the attenuation of the total $\alpha$ effect by the addition
of a term proportional to the magnetic helicity effect \citep{FB02,BB02}.

In a topological sense, magnetic helicity describes the linkage of magnetic
flux tubes \citep{Mof69}, while the kinetic helicity characterizes the
linkage of vorticity tubes.
However, there is yet another helicity, the cross helicity, that
describes the linkage of magnetic flux tubes with vortex tubes.
This quantity is important because it too is conserved by the
quadratic interactions, i.e.\ it can change only by visco-resistive
effects or by cross helicity fluxes.
Moreover, the small-scale cross helicity can itself lead to
large-scale dynamo action \citep{Yos90}.
Such a mechanism is quite different from the $\alpha$ effect, because
it corresponds to an inhomogeneous term in the dynamo equations and
could therefore play the role of a turbulent battery term.
Indeed, \cite{BU98} showed that the battery term due to cross helicity
can facilitate large-scale dynamo action in young galaxies and hence
could be responsible for the relatively strong magnetic fields
observed in such galaxies at high redshifts.

In spite of several additional studies \citep{YY93,Yok96,BC97}, 
large-scale dynamo action due to cross helicity has not 
received much attention because this effect was never seen in
simulations, nor was it found to be responsible for driving large-scale
magnetic fields found therein.
Such an effect would require that the small-scale magnetic field is
systematically aligned with the flow, i.e.\ it is either mostly parallel
or mostly anti-parallel to the flow.
Such circumstances are known to prevail in the solar wind, but here
the field comes presumably directly from the Sun and would therefore
not be produced by a dynamo.

In the present paper we consider the so-called \cite{Arc00} dynamo
\citep[see also][]{DA04,CG06} which is
driven by a forcing function that is based on the
Arnold--Beltrami--Childress (or ABC) flow, but with the cosine terms
being omitted.
This flow was first proposed by \cite{GP92} to study fast dynamo action
by calculating growth rates for the kinematic version of this flow.
The ABC flow is helical and produces efficient dynamo action \citep{GF86}.
However, the omission of cosine terms renders the flow nonhelical, so 
that there is no $\alpha$ effect, but numerical studies \citep{DA04} have shown
that such a dynamo produces magnetic fields that are
either aligned or anti-aligned with the flow almost everywhere.
This means that there is cross-helicity in the system, which can give
rise to the Yoshizawa effect and produce large-scale dynamo action.
Furthermore, owing to the conservation property of cross helicity,
such dynamos may be controlled by this effect and may also show slow
saturation behavior.
It is therefore of interest to investigate whether the formulation
for the slow saturation of $\alpha$ effect dynamos carries over to the
present case.

We begin by explaining first the simulations, discuss the features of 
the kinematic growth phase of the dynamo, and then consider the
slow saturation regime using a nonlinear dynamical feedback formalism
that is analogous to the dynamical quenching formalism for the
$\alpha$ effect. Next we argue that the kinematic growth in such
a dynamo is indeed due to the cross helicity effect. We show that
the estimated growth rate obtained from a simple model involving
the induction and momentum equations along with the evolution equation
for the small-scale cross helicity can be brought in good agreement
with our simulation results.

\section{Basic equations}

We consider here a model that is similar to that of \cite{Arc00} and
\cite{DA04} who assumed a compressible gas with an energy equation included.
However, in their model the temperature was kept approximately constant
by applying a heating and cooling term.
Here we assume instead an isothermal equation of state, i.e.\ the
pressure is given by $p=\rho\cs^2$, where $\rho$ is the density and
$\cs$ is the isothermal sound speed.
The evolution equations for the density $\rho$, velocity $\UU$, and magnetic
vector potential $\AAA$ are then
\EQ
{\DD\ln\rho\over\DD t}=-\nab\cdot\UU,
\EN
\EQ
{\DD\UU\over\DD t}=-\cs^2\nab\ln\rho+\FF
+{1\over\rho}\left[\JJ\times\BB+\nab\cdot(2\rho\nu\SSSS)\right],
\label{dUdt}
\EN
\EQ
{\partial\AAA\over\partial t}=\UU\times\BB+\eta\nabla^2\AAA,
\EN
where $\DD/\DD t=\partial/\partial t+\UU\cdot\nab$ is the advective
derivative, $\BB=\nab\times\AAA$ is the magnetic field,
$\JJ=\nab\times\BB/\mu_0$ is the current density,
$\mu_0$ is the vacuum permeability,
$\eta$ is the magnetic diffusivity, which is assumed constant,
$\nu$ is the kinematic viscosity,
\EQ
\SSS_{ij}=\half(U_{i,j}+U_{j,i})-\onethird\delta_{ij}\nab\cdot\UU
\EN
is the traceless rate of strain tensor, and
\EQ
\FF=F_0\;(\sin k_0z, \sin k_0x, \sin k_0y)
\EN
is the forcing function
where $F_0$ is an amplitude factor and $k_0$ is a wavenumber.

For analytic considerations we consider the flow to be
incompressible, i.e.\ $\nab\cdot\UU=0$ and $\rho=\rho_0=\const$.
While this simplifies the treatment significantly, it should be
remembered that the differences between compressible and incompressible
cases are not critical if the Mach number is small \citep{CG06}.
In the present paper we consider cases where the Mach number is around
0.03 (see below).
In order to simplify the notation we use units where
\EQ
k_0=\cs=\rho_0=\mu_0=1,
\EN
although in several places we shall keep these units for clarity.

The simulations have been performed using the \textsc{Pencil Code}
\footnote{\url{http://pencil-code.googlecode.com}}.
Triply periodic boundary conditions are employed for all
variables over a cubic domain of size $L\times L\times L$.
As initial condition we use zero velocity, constant density
given by $\rho=\rho_0$, and a spatially random vector
potential of sufficiently low amplitude so as to obtain
a clear initial exponential growth phase over several orders of magnitude
before nonlinear effects become important and lead to saturation of the
magnetic field.

Our simulations are characterized by the values of the magnetic
Reynolds and Prandtl numbers,
\EQ
\Rm={u_0\over\eta k_0},
\quad\mbox{and}\quad\Pm={\nu\over\eta},
\EN
respectively.
Here, we have defined $u_0=(F_0/k_0)^{1/2}$ as our reference velocity.
Occasionally we also use the visco-resistive Reynolds number,
\EQ
R_\mu={u_0\over\mu k_0}={\Rm\over1+\Pm},
\EN
where $\mu=\nu+\eta$.
Throughout this paper we restrict ourselves to the case $\Pm=1$.
The forcing amplitude is chosen such that the Mach number,
$\Ma=u_0/\cs$ is small (about 0.03), so the flow stays close to
incompressible.

The flow is of course isotropic with respect to the three coordinate
directions, so there is no preferred definition for the mean field
in this case.
Indeed, there are three equivalent definitions of two-dimensional averages
($xy$, $yz$, and $xz$ averages).
They all would lead to finite mean flows and mean magnetic fields.
In the following we consider mean fields defined by averaging over
the $x$ and $y$ directions, i.e.\
\EQ
\meanBB(z,t)={1\over L^2}\int\BB\,\dd x\,\dd y.
\EN
Throughout this paper we focus on the case $L=L_0$,
where we have defined $L_0=2\pi/k_0$.
However, on one occasion we compare with the cases $L=2L_0$ and $4L_0$,
where the domain is big enough to allow for a field configuration
that is four times bigger than the wavelength of the sine waves.
The residual, $\bb=\BB-\meanBB$, is normally referred to as the small-scale
or fluctuating field, but in the present case such a characterization
might be misleading,
because such a field is quite regular and not actually fluctuating
in the real sense of the word.
Note in particular that the forcing function has a finite average, i.e.\
\EQ
\meanFF(z)=F_0(\sin k_0z, 0, 0),
\EN
so the residual is $\ff=F_0(0, \sin k_0x, \sin k_0y)$.
It turns out that also $\meanUU$ and $\meanBB$ point mainly in the
$x$ direction.
Throughout this paper we denote the residuals by lower case characters.

\section{Simulation results}
\label{SimulationResults}

\begin{figure}\begin{center}
\includegraphics[width=\columnwidth]{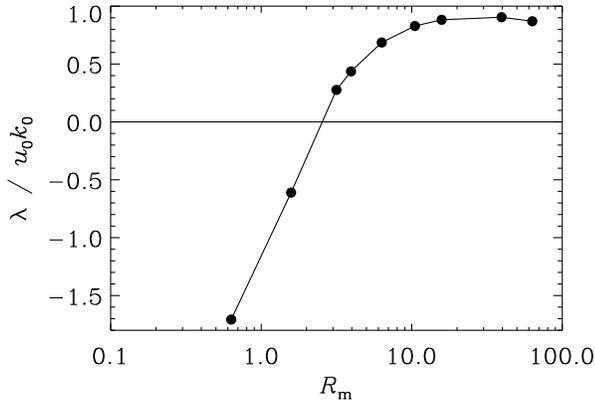}
\end{center}\caption[]{
Dependence of the dynamo growth rate $\lambda$ on $\Rm$.
Note that the critical value of $\Rm$ for dynamo action is around 3.
For larger values of $\Rm$ the growth rate levels off at a value
around $u_0k_0$.
}\label{GRm}\end{figure}

Dynamo action is possible once the value of $\Rm$ exceeds a certain
critical value of around 3; see \Fig{GRm}.
A similar curve was first shown by \cite{GP92} for the case
of a prescribed flow $U=u_0$.
For smaller values of $\Rm$ the growth rate is negative while for
larger values it levels off at a value comparable to $u_0k_0$.

\begin{figure}\begin{center}
\includegraphics[width=\columnwidth]{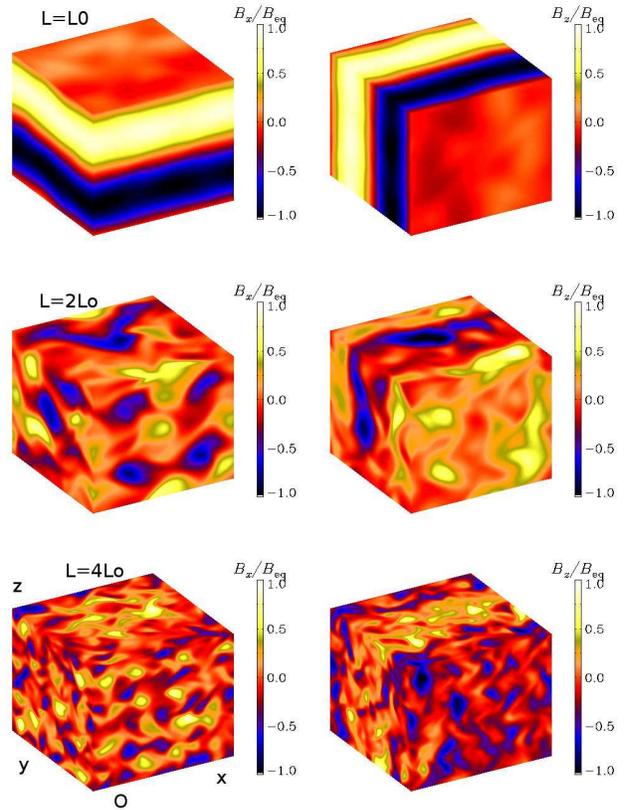}
\end{center}\caption[]{
Visualization of $B_x$ and $B_z$ on the periphery of the domain in
models with $L/L_0=1$ (upper row), 2 (middle row), and 4 (lower row)
for $\Rm=13$.
In the case $L=L_0$ the magnetic field has a scale that
is equal to that of the flow, but in the other cases
the field breaks up into smaller scale contributions with a
modulation in the $y$ direction on the scale of the domain.
The coordinate directions are indicated in the lower left panel
and the origin is indicated by O.
}\label{all}\end{figure}

One may expect the Archontis flow to be a small-scale
dynamo, which means that the scale of the field would not exceed the
scale of the flow, $L_0$.
In order to check whether this flow can also generate fields on a
scale larger than that of the flow we consider now also cases with
$L=2L_0$ and $4L_0$.
In \Fig{all} we compare visualizations of $B_x$ and $B_z$ for
$L/L_0=1$, 2, and 4.
In the case $L=L_0$ the magnetic field has a scale that
is equal to that of the flow, but in the other cases
the field breaks up into smaller scale contributions with a
modulation in the $y$ direction on the scale of the domain.
In the latter case, the field on the scale of the domain is
reminiscent to that found in helical turbulence \citep{B01},
but it is less dominant and less persistent than for $L=L_0$.
This is mainly explained by a strong reduction of net cross helicity
when the field breaks up into smaller-scale contributions.
For these reasons we focus in the remainder of this paper on
the case $L=L_0$, which is perhaps the simplest case known to produce
net cross helicity.

Those dynamos produce large-scale fields, but they are not as prominent
and persistent as in the case of large-scale dynamos that are driven by
kinetic helicity.
This is mainly because in the simulations with larger domains the
cross helicity is strongly reduced once the magnetic field breaks up
into smaller-scale fields.

\begin{figure}\begin{center}
\includegraphics[width=\columnwidth]{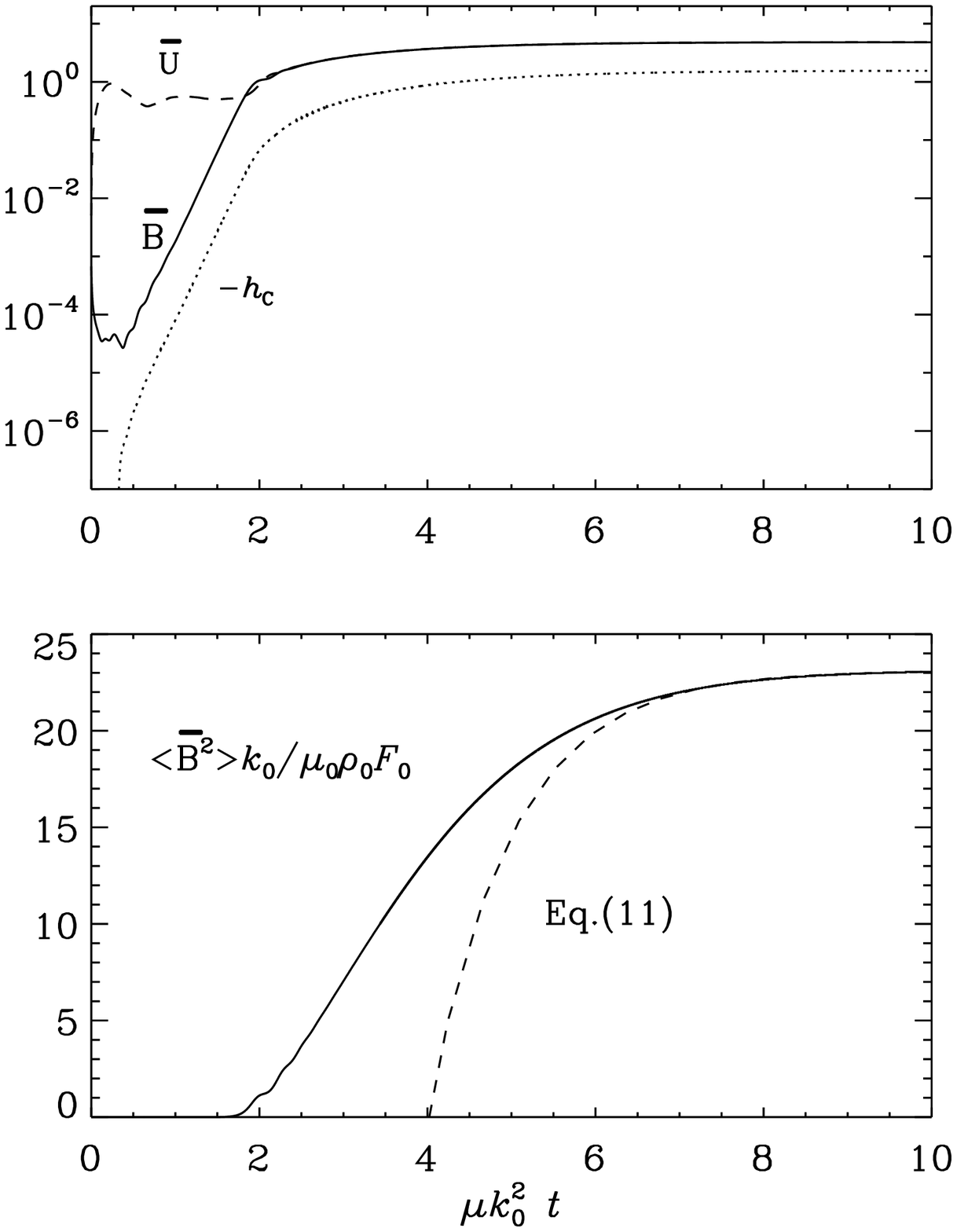}
\end{center}\caption[]{
Saturation behavior for a run with $\Rm=16$.
The dotted line shows that the simple-minded helicity constraint
formula does not describe the saturation of $\meanBB^2$ correctly.
The labels $\overline{\rm U}$, $\overline{\rm B}$, and $h_{\rm C}$
denote $\bra{\meanUU^2 k_0/F_0}^{1/2}$,
$\bra{\meanBB^2 k_0/\mu_0\rho_0 F_0}^{1/2}$,
and $\bra{\uu\cdot\bb}$ $(k_0/F_0)^{1/2}(\mu_0\rho_0)^{-1/4}$.
}\label{psat_n64k}\end{figure}

In \Fig{psat_n64k} we show the evolution of the mean magnetic field,
mean velocity and the small-scale cross helicity, $h_c=\bra{\uu\cdot\bb}$,
for a run with $\Rm=16$ in a logarithmic scale; see Panel 1 and also
the evolution of the magnetic energy compared to that of an $\alpha^2$ dynamo
on a linear scale in panel 2.
Time is normalized with respect to the microscopic visco-resistive time scale,
$(\mu k_0^2)^{-1}$.
Given that the initial magnetic field is spatially random, it is first
smoothened by resistive effects, leading to a short period where the
magnetic energy decreases.
Exponential growth occurs after about half a visco-resistive time,
and then turns into a slow saturation phase after about two
visco-resistive times, which is best seen on a linear scale
(lower panel of \Fig{psat_n64k}).
However, the late saturation behavior deviates from that of the
$\alpha^2$ dynamo, where the late evolution of the mean field is
well described by a switch-on curve of the form
\EQ
\meanBB^2\sim1-\exp(-\Delta t/\tau_\eta),
\EN
where $\Delta t=t-t_{\rm s}$ is the
time after the end of the exponential growth phase at $t=t_{\rm s}$
and $\tau_\eta=(2\eta k_1^2)^{-1}$ is the large-scale resistive time
based on the wavenumber $k_1$, which would be equal to $k_0$ in the
present case.

\begin{figure}\begin{center}
\includegraphics[width=\columnwidth]{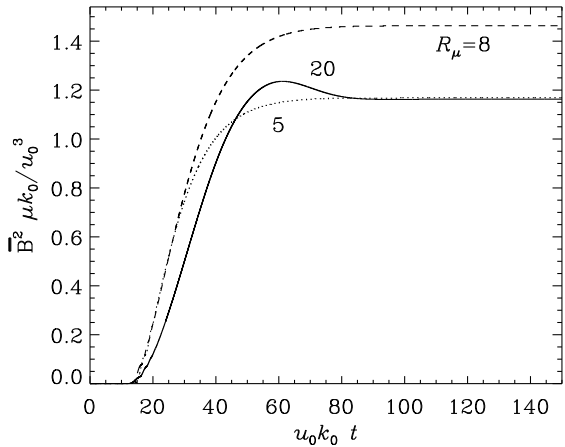}
\end{center}\caption[]{
Comparison of the saturation behavior of for three different values of
$R_\mu$.
}\label{psat_comp}\end{figure}

In \Fig{psat_comp} we demonstrate that the saturation time is
essentially independent of the value of $R_\mu$.
Here, time is expressed in dynamical units by normalizing it in terms of the
turnover time $(u_0k_0)^{-1}$.
The amplitude of the mean field increases mildly with $R_\mu$.
A suitable non-dimensional representation of the mean field is the
quantity $\meanBB^2\mu k_0/\mu_0\rho_0 u_0^2$.
This number turns out to be of order unity and only weakly dependent
on the value of $R_\mu$ for values between 5 and 20.

In the three cases displayed in \Fig{psat_comp} we have verified that
the choice of averaging is unimportant.
In other words, the results for $yz$ and $xz$ averages agree with those
for the $xy$ averages shown in \Fig{psat_comp} within 0.1--0.5 per cent.

\section{Turbulent magnetic diffusivity}
\label{Turbulent}

In a number of circumstances it has been possible to characterize
the production of mean magnetic field in terms of $\alpha$ effect and
turbulent magnetic diffusivity.
Here, ``turbulent'' refers to the commonly used name for transport
coefficients describing the evolution of mean fields rather than a
distinction between turbulent versus laminar flow properties.
Both $\alpha$ effect and turbulent magnetic diffusivity have been
determined also for other laminar flows such as the Roberts flow
\citep{Betal08b}.
However, such a description may not be applicable in the present case
because of the possible presence of the additional Yoshizawa effect.
Ignoring this complication for a moment, we can
determine the $\alpha_{ij}$ and $\eta_{ij}$ tensors in the relation
\EQ
(\overline{\uu\times\bb})_i=\alpha_{ij}\meanB_j-\eta_{ij}\meanJ_j
\label{uxb}
\EN
using the test-field method \citep{S05,S07}.
In this approach one solves an additional set of three-dimensional
partial differential equations for vector fields $\bb^{pq}$, where the labels
$p=1,2$ and $q=1,2$ correspond to different pre-determined one-dimensional
test fields $\meanBB^{pq}$.
This leads to four vector equations for $\overline{\uu\times\bb^{pq}}$
that allow us to determine all components of $\alpha_{ij}$ and $\eta_{ij}$
as functions of $z$ and $t$.
Owing to homogeneity and stationarity, it makes sense to present their
averages over $z$ and $t$.
The test-field method has been criticized by \cite{CH09} on the grounds
that the small-scale dynamo action would affect the results.
However, for magnetic Reynolds numbers of up to about 100 the results
of the test-field method have been proven to be consistent with results
from direct simulations \citep{MKTB09}.

The evolution equations for $\bb^{pq}$ are derived by subtracting the
mean-field evolution equation from the evolution equation for $\BB$.
These equations are distinct from the original induction equation
in that the curl of the resulting mean electromotive force is subtracted.
This method has been successfully applied to the kinematic case of weak
magnetic fields in the presence of homogeneous turbulence either without
shear \citep{Sur_etal08,Betal08b} or with shear \citep{B05,Betal08},
as well as to the non-kinematic case with equipartition-strength
dynamo-generated magnetic fields \citep{Betal08c,TB08}.

\begin{figure}\begin{center}
\includegraphics[width=\columnwidth]{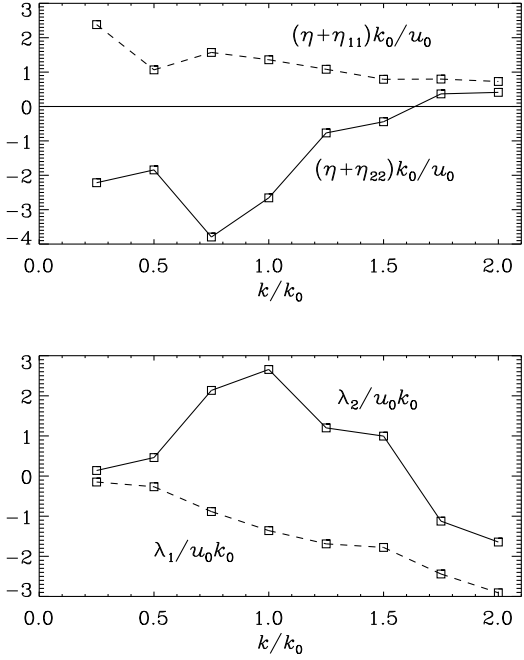}
\end{center}\caption[]{
Dependence of the normalized diagonal components of the
turbulent resistivity tensor for $\Rm=6$ (upper panel)
together with the corresponding growth rates (lower panel).
}\label{pk32}\end{figure}

Using this method,
it turns out that all components of $\alpha_{ij}$ vanish within error bars,
and that $\eta_{ij}$ has only diagonal components.
However, as shown in \Fig{pk32}, the $\eta_{22}$ component can be negative
within a limited range of wavenumbers.
(The fact that $\eta_{11}\neq\eta_{22}$ is not a priori surprising,
because both $\meanUU$ and $\meanBB$ have only components in the $x$
direction.)
One of the two growth rates,
\EQ
\lambda_1=-(\eta+\eta_{11})k_0^2,\quad
\lambda_2=-(\eta+\eta_{22})k_0^2
\EN
is therefore positive.
This suggests that there is the possibility of driving a dynamo by a
negative turbulent resistivity effect \citep{ZPF01,Urp02}.
In such a case it is important to determine the wavenumber where the
growth rate is largest.
In our case, this happens for $k\approx k_1$
(see lower of \Fig{pk32}).

In the following we discard the possibility of dynamo activity
driven through a negative turbulent resistivity effect, because
the test-field method ignores the presence of the Yoshizawa effect.
Thus, we argue that \Eq{uxb} is an {\it inadequate} ansatz that results
in an apparent negative turbulent resistivity component.
In the absence of a proper method for determining $\eta_{ij}$ we
consider now a phenomenological description of the Yoshizawa effect
using an isotropic turbulent resistivity, $\etat$.

\section{Phenomenology}

The slow saturation process found here is reminiscent of the slow
saturation process found for the $\alpha^2$ dynamo, where net magnetic
helicity is being produced on a resistive time scale.
In the present case the magnetic helicity is essentially zero,
but net cross helicity is being produced.
Owing to the conservation of cross helicity, there is the possibility
here too that full saturation requires a visco-resistive time scale,
$\tau_\mu=(\mu\kf^2)^{-1}$, where $\mu=\nu+\eta$ and $\kf$ is the
wavenumber corresponding to the typical scale of $\uu$ and $\bb$.
In our case, these fields depend essentially only on the
$x$ and $y$ directions, so $\kf^2=2k_0^2$.
The form of this relation is not known, although it is already clear
that it is not the same as in the case of the nonlinear $\alpha$ effect.
Most importantly, the saturation time does not seem to depend sensitively
on the value of $R_\mu$ (\Fig{psat_comp}).
Moreover, owing to the presence of a
forcing term in the momentum equation, the cross helicity is not
necessarily conserved in the limit $\mu\to0$, but it may change.
Indeed, under the assumption of incompressibility, the evolution of the
cross helicity per unit volume, $\bra{\UU\cdot\BB}$, is given by
\EQ
{\dd\over\dd t}\bra{\UU\cdot\BB}=\bra{\FF\cdot\BB}-\mu\bra{\WW\cdot\JJ}.
\label{dUBdt}
\EN
Here, angular brackets denote volume averages
and $\WW=\nab\times\UU$ is the vorticity.
Note the presence of the forcing term that can lead to the production of net
cross helicity if the field has a component that is aligned with the forcing.

Next, we restrict ourselves to horizontal averages, denoted by an overbar,
and consider first their evolution equations,
\EQ
{\partial\over\partial t}\meanAA=\meanUU\times\meanBB+\meanEMF-\eta\meanJJ,
\label{meanAA}
\EN
\EQ
{\partial\over\partial t}\meanUU=\meanUU\times\meanWW
+\meanJJ\times\meanBB+\meanFF+\meanFFFF-\nu\meanQQ,
\label{meanUU}
\EN
where $\meanEMF=\overline{\uu\times\bb}$ is the mean electromotive force due
to the correlation of small-scale velocity and magnetic field correlations,
$\meanFFFF=\overline{\uu\times\ww}+\overline{\jj\times\bb}$ is the mean
force due to advection and Lorentz force of small scale contributions,
and $\meanQQ=\nab\times\meanWW$ is the curl of the vorticity.
As discussed above, lower case characters denote the residual or
``fluctuating'' components, so for example $\ww=\WW-\meanWW$ is the
residual vorticity.

We note that the $\meanUU\times\meanWW$ and $\meanJJ\times\meanBB$ terms
will be of no significance, because for our one-dimensional $z$-dependent
averages only the $x$ and $y$ components of $\meanAA$ and $\meanUU$ will
be important for the evolution of the dynamo.
We assume that $\meanEMF$ has only contributions from the \cite{Yos90}
effect and from turbulent resistivity and that $\meanFFFF$ has only a
contribution from turbulent viscosity, i.e.\
\EQ
\meanEMF=\Upsilon\meanWW-\etat\meanJJ,
\label{meanEMF}
\EN
\EQ
\meanFFFF=-\nut\meanQQ.
\label{meanFFFF}
\EN
A simplified derivation of the \cite{Yos90} effect is given in \App{CrossHel},
which shows that
\EQ
\Upsilon=\tau\overline{\uu\cdot\bb}.
\label{Upsilon}
\EN
We have chosen here the symbol $\Upsilon$ instead of Yoshizawa's original
symbol $\gamma$, because $\ggamma$ is frequently used to describe the
turbulent pumping velocity.
Furthermore, $\Upsilon$ looks similar to $\gamma$ and it also reminds
of the letter Y in Yoshizawa's name.

In addition, there is also turbulent viscosity
$\nut={2\over15}\tau\overline{\uu^2}$ and turbulent resistivity
$\etat={1\over3}\tau\overline{\uu^2}$ \citep{KRP94},
although numerical simulations suggest $\nut\approx\etat$ \citep{YBR03}.
Here, $\tau$ is a typical time scale that may be estimated in terms of
the turnover time, $\tau=(\urms k_0)^{-1}$, where $\urms=\bra{\uu^2}^{1/2}$.

Inserting \Eqs{meanEMF}{meanFFFF} into \Eqs{meanAA}{meanUU}, we derive
the following evolution equation for the cross helicities of the mean
and fluctuating fields:
\EQ
{\dd\over\dd t}\bra{\meanUU\cdot\meanBB}=\bra{\meanFF\cdot\meanBB}
+\Upsilon\bra{\meanWW^2}-\muT\bra{\meanWW\cdot\meanJJ},
\label{dUmBmdt}
\EN
\EQ
{\dd\over\dd t}\bra{\uu\cdot\bb}=\bra{\ff\cdot\bb}
-\Upsilon\bra{\meanWW^2}+\mut\bra{\meanWW\cdot\meanJJ}
-\mu\bra{\ww\cdot\jj},
\label{dubdt}
\EN
where $\mut=\nu_t + \eta_t$ is the sum of turbulent viscosity and
resistivity and $\muT=\mu_t+\mu$ is the total (turbulent and
microscopic) value.
One can easily verify that the sum of \Eqs{dUmBmdt}{dubdt} gives \Eq{dUBdt}.

\begin{figure}\begin{center}
\includegraphics[width=\columnwidth]{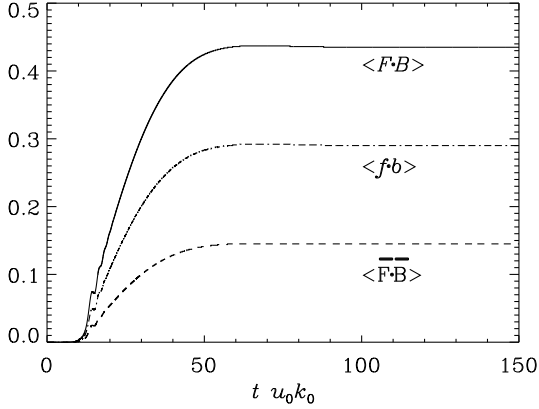}
\end{center}\caption[]{
Plot of the force-magnetic field correlations for a run with $\Rm=32$.
}\label{pfbm_n64h_fxbxm}\end{figure}

In the following we shall use \Eq{dubdt} to describe the evolution
of $\Upsilon$ fully in terms of mean field quantities.
This approach was recently perused by \cite{Kan07} for the more complete
case where kinetic and magnetic helicities are also present.
In \Eq{dubdt} the term $\bra{\uu\cdot\bb}$ is directly related
to the mean field quantity $\Upsilon$, and so is
$\bra{\ww\cdot\jj}=\kf^2\bra{\uu\cdot\bb}$.
An exception is the correlation of the forcing term with $\bb$,
i.e.\ the term $\bra{\ff\cdot\bb}$.
However, it turns out that for the Archontis flow considered here,
each of the three terms, $\bra{F_iB_i}$ for $i=1$, 2, and 3 contribute
equal amounts, so
$\bra{\meanFF\cdot\meanBB}={1\over3}\bra{\FF\cdot\BB}$ and
$\bra{\ff\cdot\bb}={2\over3}\bra{\FF\cdot\BB}$, so that we can express
\EQ
\bra{\ff\cdot\bb}=2\bra{\meanFF\cdot\meanBB}
\EN
purely in terms of mean field quantities.
The validity of these relations can be seen in \Fig{pfbm_n64h_fxbxm}
where we plot the aforementioned correlations for a run with $\Rm=32$.

\begin{figure}\begin{center}
\includegraphics[width=\columnwidth]{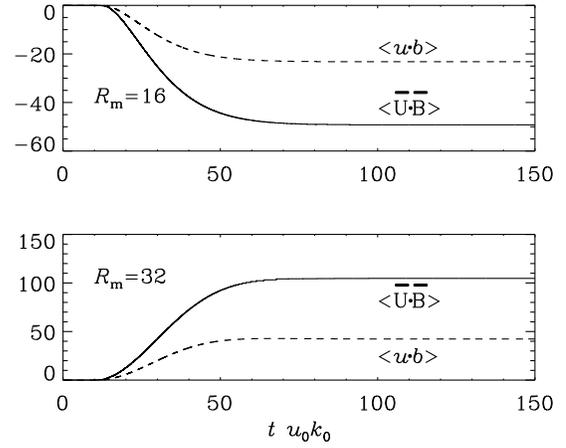}
\end{center}\caption[]{
Saturation behavior of large-scale and small-scale cross helicities
for two different values of $\Rm$.
}\label{pubm_comp}\end{figure}

With these preparations we can write down an evolution equation for
$\Upsilon$,
\EQ
{\dd\Upsilon\over\dd t}=2\tau\bra{\meanFF\cdot\meanBB}
-\tau\Upsilon\meanWW^2+\mut\tau\bra{\meanJJ\cdot\meanWW}
-\tilde{R}_\mu^{-1}{\Upsilon\over\tau},
\label{dgamdt}
\EN
where we have defined a modified visco-resistive Reynolds number
\EQ
\tilde{R}_\mu=(\mu\kf^2\tau)^{-1}.
\EN
Note that it is related to $R_\mu$ via
\EQ
\tilde{R}_\mu=(\kf/k_0)^2 (\urms/u_0) R_\mu.
\EN
Analogous to the magnetic case we can write this equation
as a quenching formula by keeping the time derivative as an implicit term,
\EQ
\Upsilon=\tilde{R}_\mu\;{2\tau^2\bra{\meanFF\cdot\meanBB}
+\tau^2\mu_{\rm t}\bra{\meanWW\cdot\meanJJ}-\tau\dd \Upsilon/\dd t
\over1+\tilde{R}_\mu\bra{\meanWW^2}\tau^2}.
\label{Upsilon}
\EN
These equations show that the generation of large-scale magnetic field by the
$\Upsilon$ term produces $\bra{\meanUU\cdot\meanBB}$ of the same sign as that
of $\Upsilon$ ($=\tau\overline{\uu\cdot\bb}$).
This is also seen in the simulations, where $\bra{\meanUU\cdot\meanBB}$ and
$\overline{\uu\cdot\bb}$ do indeed have identical signs for the same $\Rm$, 
but could individually, depending on initial conditions, have different
signs; see \Fig{pubm_comp} for two cases with different $\Rm$ values.
However, this sign property is a major difference to the case
of the $\alpha^2$ dynamo where $\bra{\meanAA\cdot\meanBB}$ and
$\bra{\aaa\cdot\bb}$ have opposite signs.
The reason for this lies in the absence of a $\Upsilon$ term that is
independent of $\overline{\uu\cdot\bb}$, i.e.\ there is only the term
$\Upsilon=\tau\overline{\uu\cdot\bb}$.
By contrast, the $\alpha$ effect has also a contribution from kinetic
helicity that is independent of magnetic helicity, i.e.\
$\alpha={1\over3}\tau\overline{\jj\cdot\bb}
-{1\over3}\tau\overline{\oo\cdot\uu}$.

It is instructive to inspect this difference by comparing \Eq{Upsilon}
with the analogous equation for $\alpha$ quenching.
Written in implicit form \citep[see, e.g.,][]{B08}, and ignoring magnetic
helicity fluxes, this equation takes the form
\EQ
\alpha={\alpha_0+\Rm\left(
\eta_{\rm t}\bra{\meanJJ\cdot\meanBB}/B_{\rm eq}^2
-\tau\dd\alpha/\dd t\right)
\over1+\Rm\bra{\meanBB^2}/B_{\rm eq}^2},
\label{QuenchExtra2}
\EN
where $B_{\rm eq}=\bra{\rho\uu^2}^{1/2}$ is the equipartition field strength
and $\alpha_0$ is the kinematic $\alpha$ effect, i.e.\ the term
proportional to $\overline{\oo\cdot\uu}$, which is the crucial term
that has no correspondence with \Eq{Upsilon}.
Another difference is the presence of the forcing term in \Eq{Upsilon}.
Apart from that the two equations are quite analogous, i.e.\
$\tilde{R}_\mu$ is replaced by $\Rm$, $\mut$ is replaced by $\etat$,
$\Upsilon$ is replaced $\alpha$, $\tau^2$ is replaced by $B_{\rm eq}^{-2}$,
and $\meanWW$ is replaced by $\meanBB$.

\section{Kinematic growth phase}
\label{Kinematic}

The equations discussed above were originally motivated by trying to
understand the late nonlinear stage of the dynamo.
However, as we see from \Eq{Upsilon}, $\Upsilon$ itself has terms proportional
to the mean field, suggesting that $\Upsilon$ should increase with the
mean magnetic field.
This is indeed the case during the kinematic stage; see the dotted line
in the upper panel of \Fig{psat_n64k}.
This suggests that the $\Upsilon$ term might also be responsible for the
kinematic exponential growth of the dynamo.
In order to identify the relative importance of this mechanism compared
with the negative magnetic diffusivity effect discussed at the end of
\Sec{SimulationResults} we investigate
a simple model based on the induction and momentum equations along with
the evolution equation for the small-scale cross helicity. The $z$-dependent
averaging procedure for the mean magnetic and velocity fields then implies,
\EQ
{\partial\meanUU\over\partial t}=\meanFF-\nuT\meanQQ,
\label{1Dmom}
\EN
\EQ
{\partial\meanBB\over\partial t}=\nabla\times(\Upsilon\meanWW-\etaT\meanJJ),
\label{1Dind}
\EN
where $\nuT=\nut+\nu$ and $\etaT=\etat+\eta$.
We write \EEqs{1Dmom}{1Dind} along with \Eq{dgamdt} in the form,
\EQ
\dot{U}=F_0-\nu_T k_0^2 U,
\label{Udot}
\EN
\EQ
\dot{B}=\Upsilon k_0^2 U-\eta_T k_0^2 B,
\label{Bdot}
\EN
\EQ
\dot\Upsilon=2\tau F_0 B+\tau k_0^2 U (\mut B-\Upsilon U)
-\tilde{R}_\mu^{-1}\tau^{-1}\Upsilon,
\label{gamdot}
\EN
where the dots denote a time derivative, and double $z$ derivatives have
been replaced by a multiplication with $-k_0^2$.
During the early kinematic phase
the mean velocity is approximately constant.
Simulation results for a run with $\Rm=32$ then yield
$\tilde{U}=U/u_0\approx0.6$.
Applying therefore \Eq{Udot} to the steady state gives $\nuT=F_0/k_0^2 U$,
i.e.\ $\nuT=1.7u_0/k_0$.

The early exponential growth of both $B$ and $\Upsilon$ is governed by
just the first terms on the r.h.s.\ of \Eqs{Bdot}{gamdot}, i.e.\
\EQ
{\dd\over\dd t}\pmatrix{B\cr\Upsilon}
=\pmatrix{
0 & k_0^2 U\cr
2\tau F_0 & 0}.
\EN
This assumes that the $\etaT$ term in \Eq{Bdot} is negligible.
Therefore the expected maximal growth rate for the Yoshizawa effect is
\EQ
\lambda_\Upsilon=\pm\sqrt{2F_0\tau k_0^2U}.
\EN
Here we may estimate $\tau$ in terms of the turnover time,
$\tau=(\urms k_0)^{-1}$.
Our dimensionless turnover time, $u_0/\urms$, is then about 0.4, so
the dimensionless growth rate is
\EQ
{\lambda_\Upsilon\over u_0 k_0}=\pm
\sqrt{2\tilde{\tau}\tilde{U}}.
\EN
This amounts to about 0.7, which is in good agreement with
the simulation data.
This suggests that the Yoshizawa effect may indeed be responsible for
driving the dynamo in the kinematic stage.

This simple model does not describe the nonlinear saturation process.
So, if one wanted to model this, one would need to assume some
{\it ad hoc} quenching prescriptions for various quantities such as
$\nut$, $\etat$, and $\tau$.
This is in stark contrast to the case of the $\alpha^2$ dynamo where
\Eq{QuenchExtra2} describes both the kinematic growth and the slow
saturation phase quite accurately in the case of periodic boundary
conditions \citep{FB02,BB02}.

\section{Conclusions}

We considered here the Archontis flow, which is a generalization of the ABC flow.
Such a flow was thought to be a small-scale dynamo capable of generating
magnetic fields at most on the scale of the flow.
However, this flow tends to produce net cross helicity, which can lead
to a mean-field dynamo effect proposed originally by \cite{Yos90}.
Direct numerical 
simulations of such flows performed with bigger box size show the presence of 
magnetic fields on scales larger than the scale of the box (\Fig{all}).
This is reminiscent of large-scale dynamos driven by kinetic helicity,
where the resulting field is however much more prominent or persistent.

The strongest cross-helicity production is found when the scale of the
domain coincides with that of the flow.
In that case dynamo action is possible once $\Rm$ exceeds
a certain critical value which in our units turns out to be $\Rm\simeq 3$. 
The present work has shown that the kinematic phase of the Archontis
dynamo can be modelled in terms of the Yoshizawa effect.
The sign of the cross helicity depends on initial conditions, so either
sign is possible for one and the same flow field.
Simple phenomenological considerations support the idea that the
Yoshizawa effect can be expressed in terms of the mean field alone;
see \Eq{Upsilon}.
This expression looks similar to the dynamical quenching formula for the
$\alpha$ effect under the constraint of magnetic helicity conservation.
However, this expression does not actually describe quenching, but growth.
So, contrary to our initial expectation, this mechanism is not constant
in time and so it does not correspond to a battery with linear growth,
as was assumed by \cite{BU98}.
Instead, it leads to exponential growth.
At the end of the exponential growth phase the dynamo shows a characteristic
saturation behavior that is reminiscent of $\alpha$ effect dynamos
that are controlled by resistive magnetic helicity evolution.
In the present case, the conservation of cross helicity was initially
thought to be responsible for this prolonged saturation behavior, but
it turns out that the presence of a forcing term in the momentum
equation can lead to a production of net cross helicity even in the
ideal limit.

It has long been speculated that the Yoshizawa effect could be relevant
in accretion discs and galaxies where differential rotation is strong
\citep{Yok96}.
However, it turns out that, unlike $\alpha$ effect dynamos that normally
have a given kinematic value of $\alpha$, the $\Upsilon$ term cannot be
calculated a priori, but it itself depends on the mean field.
The end result is again reminiscent of the $\alpha$ effect in that both
$\Upsilon\meanWW$ as well as $\alpha\meanBB$ are linear in $\meanBB$
during the kinematic growth phase.
However, the results of the test-field method show clearly that there
is no $\alpha$ effect in that case.
Indeed, the mean electromotive force has no component along the mean
magnetic field, confirming that there is no $\alpha$ effect.
There is also no shear--current (or $\meanWW\times\meanJJ$) effect
\citep{RK03,RK04}, because the off-diagonal components of $\eta_{ij}$
were found to be zero within error bars (\Sec{Turbulent}).
This supports the idea that the growth of the magnetic field is here
indeed the result of the Yoshizawa effect.
Although the $\eta_{22}$ component is found to be negative when ignoring
the Yoshizawa effect, it is argued that this result is an artifact of
using an inadequate ansatz for the mean electromotive force.

Obviously, the flow considered here is relatively simple and hardly of
direct astrophysical relevance.
However, it has been suggested that dynamos with field-aligned flows
might be particularly efficient in generating magnetic fields in the
solar tachocline \citep{Gal08}.
If those ideas can be substantiated, it would be interesting to see
whether the phenomenological description developed in the present
paper carries over also to other cases such as this tachocline model.

Another possible avenue for future research would be the study of fully
turbulent dynamos in the presence of cross helicity.
An example of this was shown in \Fig{all} where the flow was driven by
the Archontis forcing function, but on a scale that is smaller than that
of the computational domain.
Those dynamos produce large-scale fields, but they are not as prominent
and persistent as in the case of large-scale dynamos that are driven by
kinetic helicity.
This is mainly because in the simulations with larger domains the
cross helicity is strongly reduced once the magnetic field breaks up
into smaller-scale fields.

As mentioned in the introduction, the solar wind is one of the few
examples where the turbulence is believed to have net cross helicity,
but with opposite signs in the two hemispheres.
Although the solar wind is not normally thought to harbor dynamos,
there is the problem of an unexplained contribution to energy deposition
away from the source.
It would therefore be worthwhile exploring the role of the Yoshizawa
effect in the conversion of energy in solar wind turbulence.

\section*{Acknowledgments}
We thank the organizers of the program ``Magnetohydrodynamics of Stellar
Interiors'' at the Isaac Newton Institute in Cambridge (UK) for creating
a stimulating environment that led to the present work,
and Robert Cameron and David Galloway for inspiring discussions.
SS thanks Nordita for hospitality during the course of this work
and the Council of Scientific and Industrial Research,
India for providing financial support.
We acknowledge the use of the HPC facility (Cetus cluster) at IUCAA,
the National Supercomputer Centre in Link\"oping and the Center for
Parallel Computers at the Royal Institute of Technology in Sweden.
This work was supported in part by the Swedish Research Council,
grant 621-2007-4064, and the European Research Council under the
AstroDyn Research Project 227952.


\appendix

\section{Cross-helicity effect}
\label{CrossHel}

We present here a simplified derivation of \Eq{Upsilon}
using the minimal $\tau$ approximation.
We use the linearized evolution equations for the fluctuations
$\bb$ and $\uu$ and calculate
\EQ
\partial\meanEMF/\partial t=\overline{\uu\times\dot{\bb}}+\overline{\dot{\uu}\times\bb}.
\EN
In order to highlight the essence of the \cite{Yos90} term we isolate
from the very beginning the terms that are proportional to the mean vorticity.
Thus, we consider in the evolution equations of $\dot{\bb}$ and $\dot{\uu}$
only those terms that contribute to terms proportional to $\meanWW$ and write
\EQ
\dot{\bb}=+\bb\cdot\nab\meanUU+...=-\half\bb\times\meanWW+...,
\EN
\EQ
\dot{\uu}=-\uu\cdot\nab\meanUU+...=+\half\uu\times\meanWW+...,
\EN
where we have included only the antisymmetric contribution to
$\nab\meanUU$ that leads to terms with $\meanWW$, i.e.\
$\meanU_{i,j}=-\half\epsilon_{ijk}\meanW_k+$ the symmetric part,
where a comma denotes a partial derivative.
Next, we calculate $\partial\meanEMF/\partial t$ and include only terms
proportional to $\uu\cdot\bb$ by assuming
$u_ib_j=\onethird\delta_{ij}\uu\cdot\bb+$ terms proportional to
$\uu\times\bb$, but those would later not contribute to the component
of $\meanEMF$ that is parallel to $\meanWW$.
In this way we obtain from $\overline{\uu\times\dot{\bb}}$ and
$\overline{\dot{\uu}\times\bb}$ each the term
$\onethird\overline{\uu\cdot\bb}$, so
\EQ
\partial\meanEMF/\partial t=\twothird\overline{\uu\cdot\bb}+...
-\mbox{triple correlations}.
\EN
In the spirit of the minimal $\tau$ approximation we approximate the
triple correlations by a quadratic correlation in the form of a damping term,
i.e.\ we assume that the triple correlations are equal to $\meanEMF/\tau$.
Finally, assuming stationarity, we drop the time derivative and obtain
$\meanEMF=\twothird\tau\overline{\uu\cdot\bb}$.

In an alternative derivation one can write
$\UU\cdot\nab\UU=\UU\times\WW-\half\nab\UU^2$ and subsume
the gradient term in a generalized pressure term.
Splitting $\UU\times\WW$ into mean and fluctuating part yields then
directly a term $\uu\times\meanWW$ without the 1/2 factor.
The final result is then
\EQ
\meanEMF=\tau\overline{\uu\cdot\bb},
\EN
which is also the expression used here.

\label{lastpage}
\end{document}